\documentclass[aip,jcp,preprint]{revtex4-1}
\usepackage{graphicx}
\usepackage{amssymb}
\usepackage{hyperref}
\usepackage[]{subfigure}

\begin{document}
\title{Effects of the dipolar interaction on the equilibrium morphologies
of a single supramolecular magnetic filament in bulk}
\author{Pedro A. S\'anchez}
\affiliation{Institut f\"{u}r Computerphysik, Universit\"{a}t Stuttgart, D-70569 Stuttgart, Germany.}
\email{psanchez@icp.uni-stuttgart.de}
\author{Joan J. Cerd\`a}
\affiliation{Instituto de F\'{\i}sica Interdisciplinar y Sistemas Complejos, IFISC (CSIC-UIB). Universitat de les Illes Balears. E-07122 Palma de Mallorca, Spain.}
\email{joan@ifisc.uib-csic.es}
\author{Tom\'as Sintes}
\affiliation{Instituto de F\'{\i}sica Interdisciplinar y Sistemas Complejos, IFISC (CSIC-UIB). Universitat de les Illes Balears. E-07122 Palma de Mallorca, Spain.}
\author{Christian Holm}
\affiliation{Institut f\"{u}r Computerphysik, Universit\"{a}t Stuttgart, D-70569 Stuttgart, Germany.}
\email{holm@icp.uni-stuttgart.de}

\begin{abstract} We study the equilibrium morphologies of a
  single supramolecular magnetic filament in a three-dimensional
  system as a function of the effective strength of the
  magnetic dipolar interactions. The study is performed by means of Langevin
  dynamics simulations with a bead-spring chain model of freely
  rotating dipoles. We demonstrate the existence of three
  structural regimes as the value of the dipolar coupling parameter is
  increased: a coil compaction regime, a coil expansion regime and a
  closed chain regime in which the structures tend progressively to an
  ideal ring configuration. We discuss the governing effects
  of each regime, the structural transition between open and closed
  morphologies, and the reasons why we see no multiloop configurations
  that have been observed in two-dimensional systems under similar conditions. 
\end{abstract}

\maketitle

\section{Introduction}
Supramolecular magnetic filaments are assemblies of nanometer sized
magnetic particles linked by polymers or other molecules to form a
chain. These systems are the basis of a novel nanotechnology that
combines in a unique way the interesting physical properties of
magnetic nanoparticles (MNPs) with the intrinsic anisotropies of
one-dimensional semiflexible chains \cite{wang11c}.

Research on MNPs has become an extremely active field in recent years
\cite{kodama99a, gubin05a, majetich06a, morup11a}. In particular, the
self-assembly of MNPs into stable structures, mainly governed by
dipolar interactions, is a topic of great interest. The zero-field
self-assembly of magnetic colloids into dipolar rings ---with the
individual magnetic moments disposed in a head-to-tail arrangement
along the chain--- was predicted more than four decades ago
\cite{gennes70a} and since then has been extensively studied by means
of theoretical models and simulations \cite{osipov96a, morimoto03a,
 morozov03a, morozov04a, cerda08a, kantorovich08a, prokopieva09a,
  prokopyeva11a}. However, the first experimental observations of
self-assembled chains of MNPs are more recent and, in late years, have led to the finding of
different self-assembled structures of two-dimensional dispersions of
dipolar chains, including flux closure structures like rings,
necklaces or other two-dimensional multiloop configurations
\cite{butter03a, tripp03a,
 klokkenburg04a, xiong07a, wang10c, yoon10a, wei11a, benkoski11a,
 ding12a, szyndler12a}. The direct observation of closed ring-like
structures in three-dimensional dispersions of MNPs is experimentally
challenging, but their existence has been predicted theoretically by
means of different simulation models \cite{satoh96a, aoshima04a,
 yoon10a, rovigatti12a}.

On the other hand, research on magnetic filaments is rather scarce.
After the seminal works reported in the
late 1990s for the assembly of micrometer sized particles
\cite{furst98a, furst99a}, the progress in synthesis methods to
produce them has allowed the preparation of magnetic filaments made of
particles of very different materials and characteristic sizes
\cite{dreyfus05a, tabata03a, evans07a, keng07a, bowles07a,
 benkoski08a, zhou09a, benkoski10a}. In addition, the filaments can
now be made much more flexible, a property which is mainly determined
by the nature of the molecular links between the MNPs and, to some
extent, by the intensity of their magnetic interactions. So far, most
studies on single or highly diluted magnetic filaments in bulk have
been restricted to the understanding of their dynamic properties under
the action of external magnetic fields, paying a special attention to
their application as nanofluidic propellers and actuators
\cite{cebers05a, evans07a, belovs09a, fahrni09a, benkoski10a,
 benkoski11a, babataheri11a, javaitis11a, breidenich12a}. The
equilibrium configurational properties of the filaments have been
generally disregarded. Nevertheless, the chain morphology has been
pointed out as a relevant factor for some properties of the filaments
that are of high interest like its thermal and electric conductivity
or its overall coercivity \cite{wang11c}.

As a general hypothesis, it is reasonable to expect that the
configurations observed in self-assembled chains of MNPs may appear
also as equilibrium morphologies of magnetic filaments with an
adequate degree of flexibility. Nevertheless, it is not obvious which
conditions may lead to the different equilibrium structures or what
will be the behavior of the filaments under conditions which do not
lead to the self-assembly of free particles into dipolar chains. Therefore, the chain connectivity imposed by the chemical links makes magnetic filaments a system with a structural behavior that is expected to be clearly distinguishable from the corresponding to assemblies of free particles. In addition, there is
still little knowledge on the equilibrium structures of either highly diluted
self-assembled dipolar chains and magnetic filaments in bulk.

In a recent work \cite{sanchez11a} we introduced a simple
coarse-grained simulation model for the study of the equilibrium
behavior of semiflexible magnetic filaments near an attractive flat
surface. We reported the effect of the dipolar interactions on the
adsorption transition and the existence of different equilibrium
morphologies that depend on the temperature and the magnetic dipole
strength. In particular, we found different adsorbed closed chain
structures, ranging from simple rings to more complex two-dimensional
closed configurations in which not only the ends but also some middle
points of the chain become in close contact to form a number of small
linked rings. Such simple rings and multiloop structures are very
similar to the ones found in the most recent experimental observations
of self-assembled chains of magnetic nanoparticles \cite{keng07a, kim11a}.

In the present work we explore the corresponding equilibrium morphologies of magnetic
filaments in the bulk using our previously introduced model. In
particular, we intend here to qualitatively
determine the effects of the magnetic interactions on the local and
global configurations of a fully flexible dipolar chain without the
geometrical constrains imposed by the presence of an adsorbing
surface.

The work is organized as follows: in Section \ref{sec:model} we review
the details of the proposed model and the simulation method, in
Section \ref{sec:results} we present and discuss the results of our
simulations and we end with the concluding remarks in Section
\ref{sec:conclusions}.

\section{Simulation model and method}
\label{sec:model}
In order to explore the equilibrium behavior of a magnetic filament in
bulk we have taken just the required ingredients from our previous
coarse-grained model \cite{sanchez11a}: the linking potential, the
dipolar interaction and the steric repulsion between the
beads. Therefore, in our model a magnetic filament is represented as a
bead-spring chain of $N$ identical particles carrying at their centers
a point magnetic dipole, $\vec \mu$, which can freely rotate in any
direction.

It is important to remark that in our model the free rotation of the
dipoles is just a computational simplification useful for dynamical
simulations. In experimental chains of ferromagnetic particles the
molecular links tend to prevent the rotation of the particles and,
therefore, the reorientation of their respective dipoles. The natural
disposition of such chains is a permanent head-to-tail alignment of
the dipoles with respect to the chain backbone. The head-to-tail
alignment is also found experimentally for chains of superparamagnetic
particles as a consequence of a cooperative effect: the increase of
the dipole reversal barriers in every bead led by the external field
generated by its neighbors \cite{morup10a}. Nevertheless, according to
what we observed in our previous studies with this model
\cite{sanchez11a}, we expect that a spontaneous head-to-tail alignment
will be observed for large enough values of the dipole moment without
imposing any explicit constraint to the reorientation of the
dipoles. The drawback of this approach is that we should expect the
existence of a region of relatively weak dipolar interactions---either
led by a low value of the dipole moments and/or by a high system
temperature---in which the alignment will not take place and the model
will not be representative of a ferromagnetic behavior.

The details of our model are the following: the steric repulsions
between the beads are modeled by means of a Weeks-Chandler-Andersen
potential (WCA) \cite{weeks71a}:
\begin{equation}
U_{\mathrm{{WCA}}}(r)=\left\{ \begin{array}{ll}
U_{\mathrm{{LJ}}}(r)-U_{\mathrm{{LJ}}}(r_{cut}), & r<r_{\mathrm{{cut}}}\\
0, & r\geq r_{\mathrm{{cut}}}
\end{array}\right. ,
\end{equation}
where $r$ is the distance between the centers of the interacting
beads, $U_{\mathrm{{LJ}}}(r)=4\epsilon
[\left(\sigma/r\right)^{12}-\left(\sigma/r\right)^{6} ]$ is the
standard Lennard-Jones potential, $r_{\mathrm{{cut}}}=2^{1/6}\sigma$
is the shifting parameter selected to make the potential repulsive and
$\sigma$ is the characteristic diameter of the beads. Typically, the range of experimental values of $\sigma$ varies between $10$ and $200nm$. The bonds
between adjacent beads in the chain are formed by means of a finite
extensible non linear elastic potential (FENE), defined as:
\begin{equation}
U_{\mathrm{{FENE}}}(r)=\frac{-K_{f}~r_{\mathrm{{max}}}^{2}}{2}~\ln\left[1-\left(\frac{r}{r_{\mathrm{{max}}}}\right)^{2}\right],
\end{equation}
with $K_{f}=30/\sigma^{2}$, and $r_{\mathrm{max}}=1.5\sigma$. The use
of this potential implies to have the linking springs attached to the
center of the beads, therefore neglecting the eventual bond stretching
produced by rotations of the linked particles. The long-range magnetic
interactions are represented by the conventional point dipole-dipole
potential
\begin{equation}
U_{\mathrm{{DIP}}}(\vec{r}_{ij}; \vec \mu_i, \vec \mu_j)=\frac{\vec{\mu}_{i}\cdot\vec{\mu}_{j}}{\left|\vec{r}_{ij}\right|^{3}}-\frac{3\left[\vec{\mu}_{i}\cdot\vec{r}_{ij}\right]\left[\vec{\mu}_{j}\cdot\vec{r}_{ij}\right]}{\left|\vec{r}_{ij}\right|^{5}},
\end{equation}
where $\vec r_{ij} = \vec r_i - \vec r_j$ is the displacement vector
between the centers of the beads $i$, $j$ and $\vec \mu_i$, $\vec
\mu_j$ are the dipole moments associated to each bead. For spherical particles forming a ferromagnetic monodomain $\mu$ is related to the intrinsic magnetization of the material, $M_S$, and the particle diameter, $\sigma$, as $\mu = \frac{1}{6} \pi \sigma^3 M_S$. Section I of the Supplemental Material of this article \cite{sanchez13a} includes a plot with the different potentials defined in our model.

In magnetic dipolar systems it is usual to define a dimensionless dipolar coupling
parameter, $\lambda$, to represent the effective intensity of the
dipolar interactions. For a system of identical dipolar spheres, this
parameter is commonly defined as the ratio between the optimum
magnetic energy of two dipoles---corresponding to a close contact in a
head-to-tail arrangement--- and the thermal
fluctuations: $\lambda=\mu_e^{2}/\left ( k Tb^{3} \right )$, where
$\mu_e^2$ is the experimental squared dipolar moment of the spheres,
$k$ is the Boltzmann constant, $T$ the experimental temperature and
$b$ is the characteristic separation distance between the
dipoles. Typically, $b$ is the diameter of the beads or, equivalently, the
distance between first-nearest neighbors in chain-like systems. Therefore, $b$ is expected to be similar to the particle diameter parametrized in our interaction potentials, $b \sim \sigma$. We can
take reduced units for the experimental parameters by defining $\mu^2
\equiv \mu^2_e/\epsilon_e$ and $T^* \equiv k T / \epsilon_e$, where
$\epsilon_e$ is the characteristic experimental strength of the pair
interactions in our system. Therefore, we get the following expression
for the dipolar coupling parameter:
\begin{equation}
\label{eq:lambda}
\lambda=\mu^{2}/\left ( T^* b^{3} \right ).
\end{equation}
We assume the qualitative equivalence of the variation of $\mu^2$ or
$T^*$ as a reasonable approximation for the description of the
equilibrium morphologies of a filament under not too extreme
conditions. This assumption has been proven to be useful for the
characterization of other magnetic systems like ferrofluids. Thus, we
have chosen to sample $\lambda$ by taking different values of the
squared dipolar moment within a moderate range,
$\mu^{2}\in\left[0,20\right]$, at a constant reduced temperature,
$T^*=1$. However, the dependence of $\lambda$ on $b$ prevents in our
model to sample this parameter in a straightforward way. In
particular, we have a bounded but not fixed distance between
first-nearest neighbors in the chain. We expect such distance to
depend, at least, on either the value of $\mu^2$ or the local degree
of head-to-tail alignment of the dipoles. This latter dependence makes
the exact value of $b$ unknown beforehand for any simulation we could
attempt, so we can not set $\lambda$ directly as a running dipolar
parameter. On the other hand, it is very convenient to express our
results in terms of $\lambda$ in order to facilitate the comparison
with previous studies. Thus we have taken $\mu^2$ as the running
dipolar parameter of our simulations and calculated the corresponding
values of $\lambda$ by applying Equation (\ref{eq:lambda}), with $b$
being the measured average distance between the first-nearest
neighbors in our simulated equilibrium configurations. Finally, for
each selected value of the running dipolar parameter, four different
chain lengths, $N=\lbrace 10, 25, 50, 100 \rbrace$, have been sampled.

Our simulations have been performed by means of molecular dynamics
employing a Langevin thermostat in order to implicitly include the effects of the thermal fluctuations of the solvent. According to the selected reduced units, in our dynamical simulations the reduced time, $t^*$, is related to the experimental time, $t$, by $t^*= t \sqrt{\epsilon_e/(M \sigma^2)}$. In the following we provide the parameters used in our simulations. The common details of the simulation method can be found in Section II of the Supplemental Material \cite{sanchez13a}. In general, in a dynamical simulation the values of the mass, $M$, the inertia tensor, $I$, and the translational and rotational friction constants, $\Gamma_T$ and $\Gamma_R$, are irrelevant for the results associated to an equilibrium state. For simplicity, we have taken $\sigma=1$, $\epsilon
= \epsilon_e=1$, $M=1$ and the identity matrix for the inertia tensor in order to ensure isotropic rotations. For the friction constants we have taken $\Gamma_T=1, \Gamma_R=3/4$, since these values are known to produce a fast relaxation to equilibrium for dipolar systems
\cite{kantorovich08a, cerda08a}. In order to enhance the statistics and avoid the sampling overhead associated to eventual metastable configurations---which are likely to appear at high values of $\lambda$---we have applied the replica exchange molecular dynamics method (REMD) \cite{sugita99a, mitsutake01a} to our Langevin dynamics simulations. In particular, we used $\mu^2$ as the replica parameter, with a set of 66 values in the range $\mu^2 \in [0, 20]$. These values have been selected by requiring an exchange rate above 35\% for all replicas and chain lengths. We used open boundaries and an integration time step of $\delta t = 0.001$. At each simulation cycle, we performed $3\cdot 10^6$ equilibration steps followed by $2 \cdot 10^6$ further steps for measurements. A total of 800 cycles have been performed, but only the measures collected in the last 300 cycles have been used in our statistics. As a last remark on the simulation method, the limitation of our study
to a single filament has made the direct summation the method of
choice for the calculation of the long-range dipolar interactions. The
simulations have been carried out using the coarse-grained simulation
package ESPResSo 3.0.2 \cite{limbach06a}.

\section{Results and discussion}
\label{sec:results}
In order to present our results in terms of the dipolar coupling
parameter, $\lambda$, as defined in equation \ref{eq:lambda}, the
first property to be determined from the equilibrated filament
configurations is the average distance between the first-nearest
neighbors, or mean bond length $\langle b \rangle$, as a function of
the running dipolar parameter---the squared dipole moment,
$\mu^2$---and the chain length, $N$.
\begin{figure}[b]
 \centering
 \includegraphics*[width=0.55\columnwidth]{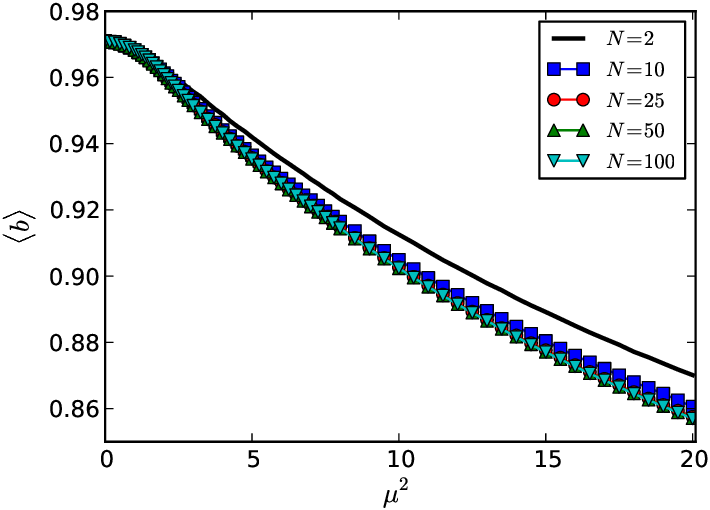}
 \caption{Mean bond length, $\langle b \rangle$, as a function of the
   squared dipole moment, $\mu^2$, for different filament
   lengths. Error bars are smaller than symbols.}
 \label{fig:bondlengths}
\end{figure}
Figure \ref{fig:bondlengths}
shows the values of $\langle b \rangle$ obtained for every selected
chain length along with the corresponding bond length of a
dimer---i.e., a chain formed by just two dipoles---which is the
minimal conceivable conformation in this system. As expected,
for all the cases studied, $\langle b \rangle$ decreases smoothly with
increasing strength of the dipole moment. On the other hand, the chain
length has a significant impact just for the shortest chains at high
dipole moments, for which the observed decrease of $\langle b \rangle$
is lower. For longer chains, the change of $\langle b \rangle$ is
almost independent of $N$. The correct interpretation of this effect requires a complete understanding of the arrangement of the dipoles in the chain as $\mu^2$, and consequently $\lambda$, are increased. Such discussion is placed in Section \ref{sec:closed}.
Finally, once the dependence of $\lambda$ on $\langle b \rangle$ has been numerically
established for each selected value of $\mu^2$ and $N$, we can express
all our results in terms of $\lambda$.

\begin{figure}[]
 \centering
 \includegraphics*[width=0.55\columnwidth]{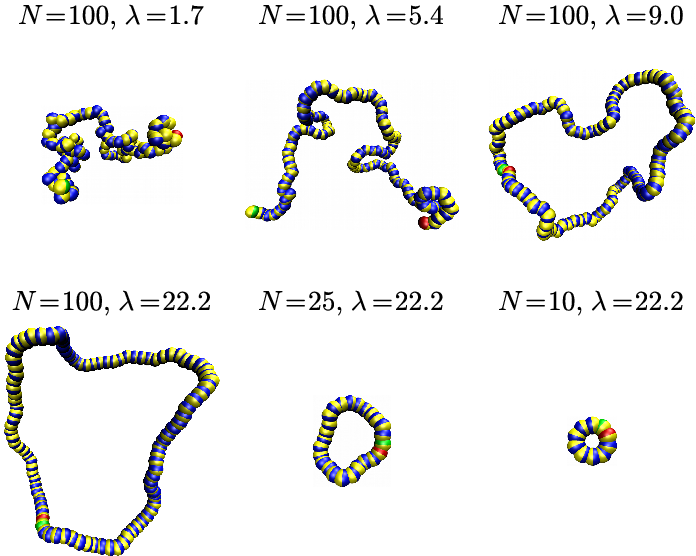}
 \caption{Selected snapshots of equilibrium configurations obtained
   for different values of the dipolar coupling parameter, $\lambda$,
   and chain lengths, $N$. The magnetic beads are represented as
   two-color spheres, with dark/light colors indicating the orientation of the
   associated dipole. The spheres at the chain ends have a different dark color to be easily identified within the closed morphologies. For the case of the lowest value of $\lambda$, the scale of a region of the chain has been increased for a better observation of the clustering and disordered orientation of the dipoles.}
 \label{fig:configs}
\end{figure}
A direct inspection of the equilibrium morphologies obtained in our
simulations confirms the expectation discussed in Section \ref{sec:model} about the head-to-tail alignment of the dipoles, as well as the existence of
important structural changes with the effective intensity of the
dipolar interactions. Figure \ref{fig:configs} illustrates these
changes for some selected values of $\lambda$ and $N$. In general, for
low values of $\lambda$ the chains show a shape similar to a swollen
random coil, with a high disorder in the orientations of the
dipoles. As $\lambda$ increases, the dipoles tend effectively to align
with the chain backbone in a head-to-tail configuration, while the
backbone becomes more straight. At even larger values of
$\lambda$, the chains adopt an irregular ring-like closed structure,
with their ends becoming permanently in close contact. Finally, these
ring-like morphologies tend to reduce their irregularities as $\lambda$
is further increased. Shorter chains apparently show less backbone
irregularities at high values of $\lambda$, getting closer to the
two-dimensional symmetry of an ideal ring. The rest of Section
\ref{sec:results} is devoted to the formal analysis of these
qualitative observations by means of different structural parameters.

\subsection{Characteristic equilibrium structures}
\label{sec:structures}
\begin{figure}[]
 \centering
 \subfigure[]{\label{fig:radiuses-mean}\includegraphics*[width=0.59\columnwidth]{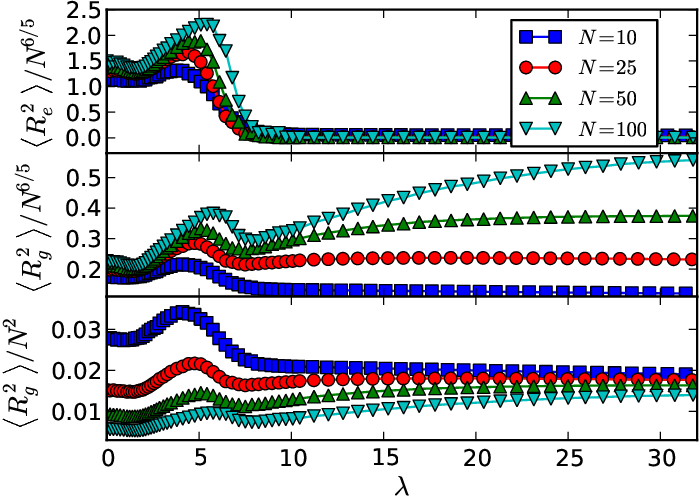}}\\
 \subfigure[]{\label{fig:radiuses-histo-Re}\includegraphics*[width=0.275\columnwidth]{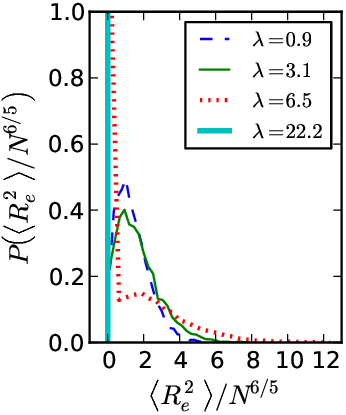}}
 \subfigure[]{\label{fig:radiuses-histo-Rg}\includegraphics*[width=0.29\columnwidth]{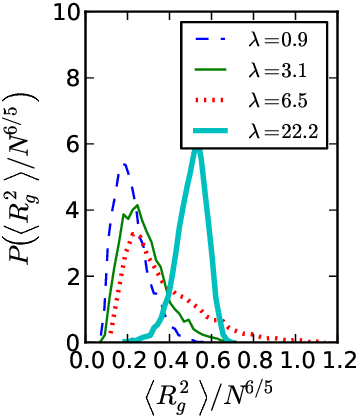}}
 \caption{(a) Scaled mean squared value of the end-to-end distance
   (upper panel) and radius of gyration (middle and lower panels), as
   a function of the dipolar coupling parameter, $\lambda$, for
   different chain lengths. The scaling shown in the upper and middle
   panels corresponds to the behavior expected for a self-avoiding
   walk, $\left \langle R_g^2 \right \rangle \propto N^{6/5}$. The
   scaling in the lower panel corresponds to the ideal ring behavior,
   $\left \langle R_g^2 \right \rangle \propto N^{2}$, represented by the solid line. (b) Probability
   histograms of $\left \langle R_e^2 \right \rangle /N^{6/5}$
   obtained for $N=100$ and some selected values of $\lambda$. (c)
   Corresponding probability histograms for $\left \langle R_g^2
   \right \rangle /N^{6/5}$.}
 \label{fig:radiuses}
\end{figure}
The squared radius of gyration, $R_g^2$, and end-to-end distance,
$R_e^2$, are useful parameters for the characterization of the global
shape of chain-like structures. Figure \ref{fig:radiuses-mean} shows
the variation of the scaled mean value of such parameters as a
function of $\lambda$. Some selected examples of their probability
distributions are also shown in Figures \ref{fig:radiuses-histo-Re}
and \ref{fig:radiuses-histo-Rg}. Two scaling
relationships $\left \langle R_g^2 \right \rangle \propto N^{2\nu}$ have
been assumed in the representation of the data in Figure
\ref{fig:radiuses-mean}: the upper and middle panels show the values
of $\langle R_e^2 \rangle$ and $\langle R_g^2 \rangle$ scaled with the
exponent $2\nu = 6/5$. This exponent corresponds to an ideal
self-avoiding walk, which is the expected behavior for this system in
the limit $\lambda \rightarrow 0$. The lower panel of Figure
\ref{fig:radiuses-mean} shows $\langle R_g^2 \rangle$ scaled with the
exponent $2\nu=2$, that corresponds to an ideal ring structure 
expected for $\lambda \gg 1$. We can observe that the
curves tend to collapse at different regions according to each scaling
behavior:  $2\nu \rightarrow 6/5$ at low $\lambda$ values and $2\nu \rightarrow 2$ at large $\lambda$.
However, it is remarkable that a perfect fit has not been reached in any case. This result indicates that
the system is far from an ideal behavior, specially in the limit of
large $\lambda$ values.


Independently of the scaling applied to the data, the results shown in
Figure \ref{fig:radiuses} clearly indicate the existence of three
different regimes and a transition-like structural change within the
explored range of parameters: for $\lambda \lesssim 2$ the chain
structures tend to compact slightly with $\lambda$, an effect that
increases with the chain length. For $2 \lesssim \lambda \lesssim 6$
this tendency is reversed, with a significant expansion of the
structures until a maximum of the overall extension is reached. At $\lambda \sim 6$ the
chain overall extension experiences a remarkable drop, with the end-to-end
distance falling to its minimum equilibrium value, corresponding to
the close contact separation between two non-bonded, head-to-tail
oriented particles. In addition, the corresponding probability
distributions show the existence of notably bigger fluctuations of
both structural parameters in this region. This important change
corresponds to the structural closure transition that leads to the
ring-like morphologies pointed above. Finally, for $\lambda \gtrsim 6$
the end-to-end distance remains in its minimum value, showing almost a
delta function in its probability distribution, and the radius of
gyration tends monotonically to a plateau. The value of this plateau
is slightly lower than the corresponding to an ideal ring formed by
$N$ beads of effective diameter $b$, $R_{g,\ ring}^2=b^2/[4
\sin^2(\pi/N)] \ \sim [N b /(2 \pi)]^2$. It is also remarkable that
the tendency to the plateau and its similarity to the value expected
for an ideal ring are significantly lower for the longest chains. As
an example, for $N=10$ and $\lambda \approx 30.39$ the mean squared
radius of gyration measured for our equilibrium structures is $\langle
R_g^2 \rangle \approx 1.92$, while an equivalent ideal ring would have
$R_{g,\ ring}^2 \approx 1.95$. On the other hand, for $N=100$ and
$\lambda \approx 30.79$ we have measured $\langle R_g^2 \rangle
\approx 139.25$ in front of the ideal value $R_{g,\ ring}^2 \approx
186.88$. Therefore, the difference between the measured and the ideal
values for $R_g^2$ at $\lambda \sim 30$ is of the order of $10\%$ for
$N=10$ and $25\%$ for $N=100$. This effect will be better illustrated
in Section \ref{sec:closed}.
\begin{figure}[!t]
 \centering
 \includegraphics*[width=0.56\columnwidth]{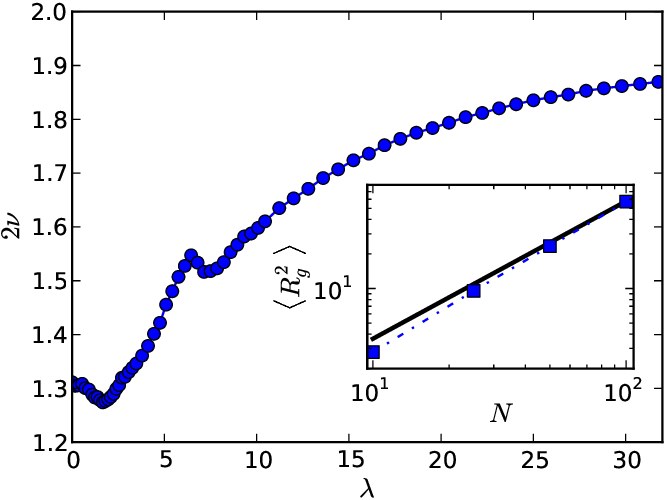}
 \caption{Dependence with $\lambda$ of the fitted exponent, $2\nu$,
   corresponding to the scaling relation for the average squared radius of
   gyration with respect to the chain length, $\left \langle R_g^2 \right \rangle \propto N^{2\nu}$
   (main plot). The inset shows how the values of $\left \langle R_g^2
   \right \rangle$ measured for $\lambda=0$ (data points) deviate from
   the ideal scaling, given by $2\nu=6/5$ (solid line).}
 \label{fig:scaling}
\end{figure}

The strong dependence of the closed structures with the chain length
discussed above explains the deviation with respect to the ideal
scaling behavior for large values of $\lambda$ that has been observed
in Figure \ref{fig:radiuses-mean}. The deviation corresponding to
small values of $\lambda$ is, instead, a consequence of being not yet in
the asymptotic scaling regime. This effect can be evidenced by fitting the scaling
exponent $2\nu$ to the available data. In Figure \ref{fig:scaling} we
have represented the fitted exponent as a function of the dipolar
coupling parameter. The fit has been performed by using all the
simulated chain lengths. The result is qualitatively consistent with
the theoretical expectations: we can observe that the value of $2\nu$
is effectively bounded by the values corresponding to the Flory
exponent for the three-dimensional swollen random coil, $2\nu=6/5$,
and the ideal ring, $2\nu=2$. The local peak observed at $\lambda \sim
6$ corresponds to the maximum expansion of the chain prior to the
closure transition. However, the inset of Figure \ref{fig:scaling}
evidences that the exponent fitted for $\lambda=0$ is slightly higher
than the theoretical value due to the fact that we have not yet reached
the asymptotic scaling regime in our sampling. An accurate estimation
of the scaling exponent for the limit $\lambda \rightarrow 0$ would
require the simulation of much longer chains.

\begin{figure}
 \centering
 \subfigure{\label{fig:histo-bondangles-N10}\includegraphics*[width=0.29\columnwidth]{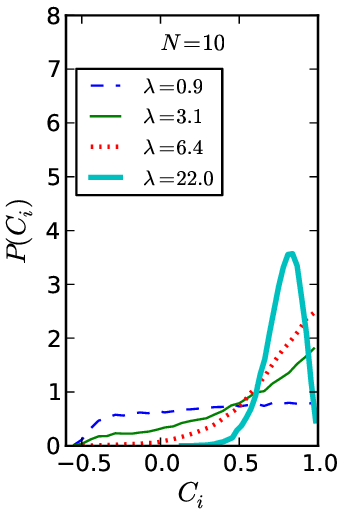}}
 \subfigure{\label{fig:histo-bondangles-N100}\includegraphics*[width=0.29\columnwidth]{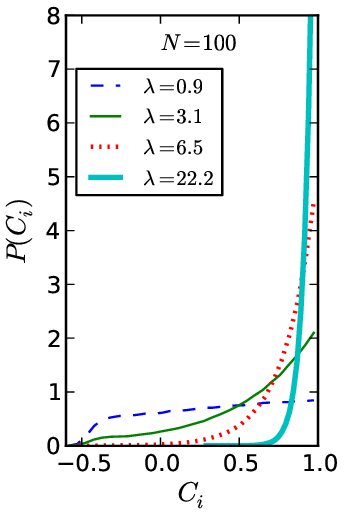}}
 \caption{Probability distributions of the cosine bond angle
   parameter corresponding to $N=10$ (left) and $N=100$ (right)
   obtained for different values of the dipolar coupling parameter.}
 \label{fig:bondangles}
\end{figure}

Another general effect of the dipolar interactions on the chains is to
make the backbone locally more straight as $\lambda$ is increased, apparently in
a rather independent way with respect to the global structure. This
magnetically driven decrease of the backbone local curvature is similar to the effect led by the chain stiffness associated to bond bending potentials
\cite{sanchez11a}. Nevertheless, it is important to keep in mind that
the dipolar interactions are not strictly local as is the case of the bond stiffness. In other chain-like systems governed by long-ranged interactions---like, for example, in
polyelectrolyte systems---the effects of the intensity of such
interactions on the local structure of the chain are usually studied
by means of the persistence length \cite{micka96a, micka97a}. In our
case, however, the existence of a closure transition led by the dipolar interactions introduces a non-local dependence of this parameter (for a discussion on the locality of the
persistence length see for instance ref. \cite{bacova12a} and
references therein). Therefore, we choose to analyze a simple but
representative local parameter: the cosine bond angle
distribution. This is defined as the probability distribution of the
vector product of adjacent unitary bond vectors for every position $i$
along the chain:
\begin{equation}
C_{i}=\hat{b}_{i-1, i}\cdot\hat{b}_{i, i+1},
\end{equation}
where $\hat{b}_{i, j}$ is the unitary vector pointing in the direction
from the center of bead $i$ to the center of bead $j$. Figure
\ref{fig:bondangles} shows the distributions for the cases $N=\lbrace
10, 100 \rbrace $ corresponding to some selected values of the dipolar
coupling parameter. The results show a continuous change with
$\lambda$ from an almost flat distribution, with a wide domain of
values for $C_i$, to a distribution that progressively approaches a
delta function at the point corresponding to an ideal ring:
\begin{equation}
\label{eq:bondcoslimit}
 C_i(\lambda\gg1) \rightarrow \delta [\cos(2\pi/N)].
\end{equation}
The almost flat distribution observed at low values of $\lambda$
indicates an insignificant correlation between adjacent bond
vectors. For high values of $\lambda$ the adjacent bond vectors are
highly correlated, as this corresponds to a locally straight
backbone. In all cases $C_{i}$ is unable to take values far below -0.5
as a direct consequence of the steric repulsions.

In Sections \ref{sec:open} and \ref{sec:closed} we analyze in more
detail the structural properties found for low and high values of
$\lambda$, respectively.

\subsection{Properties of open structures}
\label{sec:open}
\begin{figure}[!t]
 \centering
 \includegraphics*[width=0.56\columnwidth]{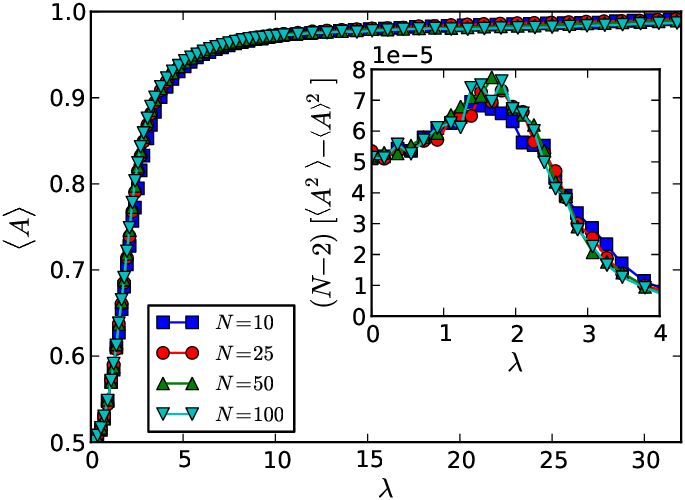}
 \caption{Average bond-dipole alignment modulus, $\langle A \rangle$,
   as a function of the dipolar coupling parameter for different chain
   lengths. Its corresponding fluctuations for low values of
   $\lambda$, scaled according to the sampling statistics, are shown
   in the inset.}
 \label{fig:bonddipole}
\end{figure}

We have shown that two different regimes are found for open
structures. In order to understand these regimes and determine at which point the model becomes representative of a ferromagnetic behavior, we have analyzed in
the first place the degree of alignment of the dipoles with the chain
backbone. This alignment can be easily calculated by means of the
bond-dipole alignment modulus, $A$, defined as \cite{sanchez11a}:
\begin{equation}
A=\frac{1}{N-2}\sum_{i=2}^{N-1}\left|\hat{b}_{i-1,i+1}\cdot\hat{\mu}_{i}\right|,
\end{equation}
where $\hat \mu_i$ is a unit vector parallel to the dipolar moment of
bead $i$ and $\hat{b}_{i-1,i+1}$ is the unit vector parallel to
the displacement vector between the centers of the beads $i-1$ and
$i+1$. We expect that for any given equilibrium configuration found in
our system this simple parameter will take values from 1/2 to 1. The
value 1/2  corresponds to a distribution of dipole orientations
completely uncorrelated with the chain backbone, whereas the value 1 is associated to a
configuration in which all dipoles are perfectly aligned with it. Figure
\ref{fig:bonddipole} shows the average and fluctuations of this
parameter as a function of $\lambda$ for every chain length. From a fully uncorrelated state at $\lambda=0$, the alignment of the dipoles grows quickly with $\lambda$ up to $\langle A \rangle \sim 0.95$ at $\lambda \sim 5$, to smoothly approach to its maximum value $\langle A \rangle =1$ afterwords. 
For all cases, the fluctuations show an absolute maximum at around $\lambda \sim 1.6$, but neither the average nor the fluctuations exhibit any significant dependence on the chain length.
Interestingly, the maximum in the fluctuations corresponds quite well
with the minima found for $R_g^2$ and $R_e^2$ that separate the two
behaviors associated to open structures. We conclude that a compaction
or an expansion of the open chains is obtained depending on the degree
of alignment of the dipoles with the chain backbone. In particular,
disordered dipoles tend to favor the chain compaction by increasing
their lateral contacts and allowing the formation of small, disordered
and weakly interacting three-dimensional clusters, like the ones that
can be observed in Figure \ref{fig:configs} for the lowest value of
$\lambda$. Such disordered aggregates should reduce the overall
extension of the chain. The monotonous decrease of the bond length
with $\lambda$ may help slightly to the overall compaction in this
region, but its impact can be neglected in front of the variations of
the lateral contacts. An estimation of the number of energetically and
entropically favorable lateral contacts, obtained by applying the
criteria established for the aggregation of magnetic particles in
ferrofluids \cite{holm06a}, supports this conclusion. The details of
such estimation can be found in Section III of the Supplemental
Material \cite{sanchez13a}.

The analysis of $\langle A \rangle$ suggests that our free rotating
dipoles tend to display a ferromagnetic behavior for $\lambda > 3$,
where the bond dipole alignment fluctuations drop almost to zero,
independently of the chain length.
This limiting value can be reasonably identified with the well known condition for the self-assembly of spherical MNPs into dipolar head-to-tail chains \cite{gennes70a, pshenichnikov00a, pshenichnikov01a, wang02a, klokkenburg04a, wang03a}. Such condition simply reflects the fact that thermal fluctuations should be less significant than the magnetic energy of two head-to-tail aligned dipoles in order to allow the formation of a stable dipolar-driven aggregate. In our case, the bonds keep the chain connectivity even at the entropically dominated region $\lambda<3$, but under such condition the dipoles are unable to remain persistently aligned and rotate like free particles. It is for $\lambda>3$ when the head-to-tail alignment becomes stable and the chains tend to be more straight as the alignment increases. In summary, the behavior of the open structures is mainly the result of the interplay between the thermal fluctuations and the dipolar interactions, both acting at a relatively short scale around each particle.

\subsection{Closure transition and properties of closed structures}
\label{sec:closed}
The inspection of the configurations and the behavior of the radius of
gyration and the end-to-end distance has evidenced the existence of a
closure transition in the bulk similar to the one observed for chains adsorbed on a flat substrate \cite{sanchez11a}. It is known that the dipolar energy
of a chain of $N>4$ dipoles disposed into a head-to-tail closed ring
is lower than the corresponding to a head-to-tail straight
arrangement of the same length \cite{jacobs55a}. In particular, for
$N>4$, the decrease of the dipolar energy led by the added
head-to-tail close contact between the chain ends overcomes the effect
of the misalignments introduced by the ring curvature. On the
other hand, a closed chain structure has a lower configurational
entropy than an open one. Therefore, the closure transition separates
the region of configurations dominated by entropy, corresponding to
the open chain structures, from the energy-dominated region of the closed
structures.

\begin{figure}[]
 \centering
 \includegraphics*[width=0.56\columnwidth]{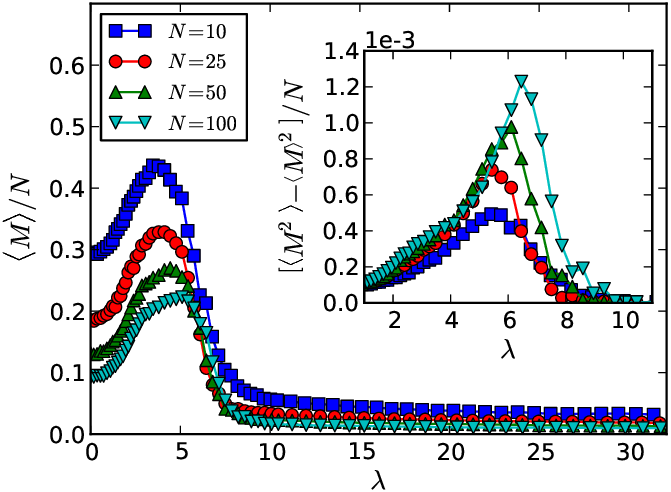}
 \caption{Mean total magnetization as a function of the dipolar
   coupling parameter for different chain lengths and its
   corresponding fluctuations (inset).}
 \label{fig:closure}
\end{figure}
In order to analyze in more detail the closure transition of our
equilibrium structures, we have chosen the total magnetization of the
chain, $M$, as the characteristic parameter. $M$ is simply defined as the
module of the sum of the unit dipolar moment vectors along the
chain:
\begin{equation}
M=\left \| \sum_{i=1}^N\hat{\mu}_{i} \right \|.
\end{equation}
The behavior of this parameter is expected to be qualitatively very
similar to the one observed for the end-to-end distance
\cite{sanchez11a}, but the accurate estimation of the fluctuations of
$M$ is found to be easier than the corresponding to $R_e^2$. Figure
\ref{fig:closure} represents the relative mean value of $M$ and its
fluctuations. Initially, $M$ grows with $\lambda$ up to a maximum
value, from which a sudden drop to almost zero is observed around
$\lambda \sim 6$. For higher values of $\lambda$ it remains
persistently close to zero. The almost zero value of $M$ indicates
that the magnetic flux along the chain is following a nearly closed
trajectory, as corresponds to a closed structure of head-to-tail
aligned dipoles. Remarkably, the fluctuations of $M$, shown in the
inset of Figure \ref{fig:closure}, display clear peaks indicating the
position of the transition points, $\lambda_0$. These transition
points shift to higher values of the dipolar coupling parameter as the
chain length increases: from $\lambda_0 \sim 5.5$ for $N=10$ to
$\lambda_0 \sim 6.5$ for $N=100$. These closure points are consistent
with the structural phase diagram known for low density dispersions of
free MNPs \cite{holm05b}. However, it is well known that
self-assembled structures of free MNPs tend to be quite heterogeneous,
with a coexistence of open and closed structures in a wide range of
parameters. This heterogeneity is not limited to experimental
observations---in which other more complex short ranged interactions,
hard to control and predict, might play an important role---but is
also found in simulations with other minimalistic models similar to
ours. Therefore, the self-assembly of free magnetic particles does not
represents a structural transition. This makes a significant
difference with our dipolar chains, for which the chain connectivity
leads to a well defined transition-like behavior.

\begin{figure}[]
 \centering
 \includegraphics*[width=0.59\columnwidth]{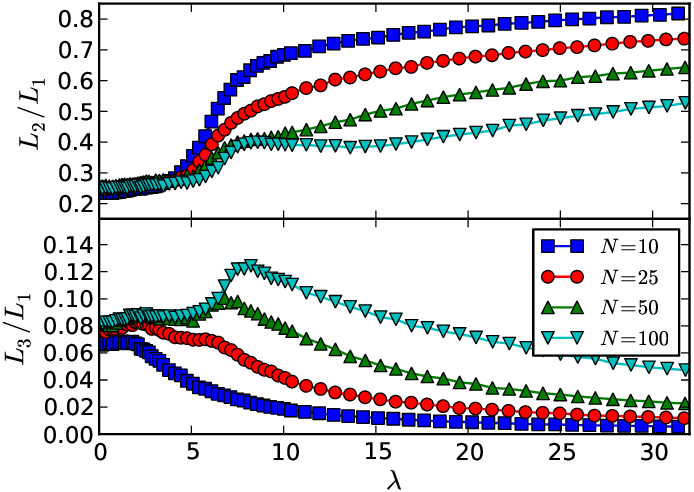}
 \caption{Shape parameters of the chains for every sampled value of
   $\lambda$ and $N$. These shape parameters are defined as the ratios
   of the average principal moments of the gyration tensor: $L_2/L_1$
   (upper panel) and $L_3/L_1$ (lower panel).}
 \label{fig:shapes}
\end{figure}

We have seen that for $\lambda > \lambda_0$ the chains remain in a closed
configuration, but it is left to determine how
far they are from an ideal ring configuration. This can
be characterized by means of different shape parameters calculated
from the average principal moments of the gyration tensor of the
sampled configurations, $L_1 \geq L_2 \geq L_3$. Here we choose to
analyze the ratios of the average second and third moments to the
first one, $L_2/L_1$ and $L_3/L_1$ respectively. Figure
\ref{fig:shapes} shows the behavior of these parameters with $\lambda$
for every chain length. For $\lambda=0$ we can observe that our
measures are in good agreement with the ratios corresponding to the
shape anisotropy of a swollen random coil \cite{solc71a}, $L_1:L_2:L_3 \approx
12:3:1$. For $\lambda \gg 1$, the shape parameters tend to $L_2/L_1 \rightarrow 1$ and $L_3/L_1 \rightarrow 0$, values that correspond to the ideal ring. This result evidences the tendency to approach an ideal ring structure as $\lambda$ increases. 
The strong dependence of this tendency on the chain length also evidences that the ideal ring is more difficult to achieve for the longest chains. 
At values of $\lambda$ in the vicinity of the closure transition point $\lambda_0$, the behaviour is rather complex and  strongly dependent on the chain length. In particular, for $\lambda \gtrsim \lambda_0$ a monotonic increase of $L_2/L_1$ can be observed for chain lengths $N < 100$. For $N=100$, instead, there exists an interval just above $\lambda_0$ in which $L_2/L_1$ slightly decreases, indicating an slight increase of the eccentricity of the closed chain in the plane of its first two principal axes. This suggest that further structural regimes with different shape anisotropies may appear as the chain length increases. An adequate characterization of such eventual new regimes will require the exploration of a wider range of chain lengths.

At the beginning of Section \ref{sec:results} we provided an evidence of a significant impact of the chain length on the average bond length, $\langle b \rangle$, mainly for short chains and strong dipolar interactions (see Figure \ref{fig:bondlengths}). This behavior can be explained by the interplay between the dipolar interaction that every particle experiences with second and further nearest neighbors in the chain and the local curvature imposed by the ring-like structures found for strong dipolar interactions: on one hand, dipolar interactions with further nearest neighbors are only significant for high dipolar moments and a good head-to-tail alignment with such close neighbors. Under these conditions, the bonding potential well between first nearest neighbors is shifted to a shorter equilibrium distance (see Figure 1 in the Supplemental Material \cite{sanchez13a}). However, only long enough chains are able to keep a high local curvature---i.e., a good alignment between further nearest neighbors---when they form a ring-like structure at high values of $\lambda$. Instead, this effect of non-first nearest neighbors is hindered by the increased misalignment led by a shorter chain ring.

Finally, and as a difference with previous results on magnetic filaments adsorbed onto a
planar surface \cite{sanchez11a}, we found no trace of two-dimensional multiloop
structures in the bulk. 
Smooth two-dimensional multiloop
structures can be slightly more energetically favorable than a single
ideal ring, provided they keep the closed head-to-tail contact between
the chain ends and add some favorable lateral contacts. On the other
hand, the formation of such lateral contacts between middle points in
the chain induces a decrease in the configurational entropy with
respect to a single ring. The balance between the changes in the
configurational energy and entropy associated to the formation of
multiloop structures depends on the system dimensionality: the
presence of a steric or adsorbing surface imposes a two-dimensional
arragement of the chains, leading to a significant decrease in their
configurational degrees of freedom. Due to this entropy reduction
mechanism, the formation of multiloop chains requires a smaller change
of entropy in two-dimensional systems than in the bulk, thus it is
favored in a broad range of parameters. Nevertheless, we expect that
multiloop configurations will be observed in bulk for low enough
temperatures, when energetic contributions overcome the entropic ones.

\section{Summary and concluding remarks}
\label{sec:conclusions}
We have studied the equilibrium behavior of a single flexible
supramolecular magnetic filament in bulk as a function of the dipolar
coupling parameter, $\lambda$, and the chain length.  The study has
been carried out by means of Langevin molecular dynamics simulations
using a bead-spring model of linked dipoles. Despite the fact that the dipoles in our
model are free to rotate in any direction, they adopt a persistent
head-to-tail alignment with the chain backbone for values of the
dipolar coupling parameter above the limiting condition for stable
self-assembly of MNPs, $\lambda>3$.

Three different structural regimes and a chain closure transition have
been found in our model: an open coil compaction regime for $\lambda
\lesssim 2$, in which the dipoles are poorly aligned and form small
aggregates of favorable lateral contacts; an open coil expansion
regime for $2 \lesssim \lambda \lesssim 6$, in which the onset of the
ferromagnetic alignment of the dipoles takes place; a magnetic flux
closure transition at $\lambda \sim 6$, signaling the separation
between the entropy and energy dominated structures; and finally, for
$\lambda \gtrsim 6$, a regime of closed ring-like structures that tend
to an ideal ring as $\lambda$ increases.

We have analyzed the local stiffening of the chain induced by the
increasing local alignment of the dipoles with $\lambda$, as well as
the significant impact of the chain length on the equilibrium behavior
of the system and its configurational entropy. We also discussed in
our three-dimensional system the absence of multiloop structures like
the ones found theoretically and experimentally in two-dimensional
systems. We attributed this fact to the excess of configurational
entropy in the bulk within the range of parameters explored.

Our results are found to be consistent with the known properties of
self-assemblies of free MNPs. However, we have shown that the chain
connectivity makes a difference with respect to dispersions and
assemblies of free magnetic particles by imposing a more coherent and
predictable behavior. This suggests that magnetic filaments can have
advantages as building blocks for the synthesis of complex magnetic
nanostructures.

\section*{Acknowledgments}
We thank Peter Ko\v{s}ovan for useful discussions and the
following organizations for providing the necessary computational resources: bwGRiD
\footnote{BwGRiD (http://www.bw-grid.de), member of the German D-Grid
  initiative, funded by the Ministry for Education and Research
  (Bundesministerium fuer Bildung und Forschung) and the Ministry for
  Science, Research and Arts Baden-Wuerttemberg (Ministerium fuer
  Wissenschaft, Forschung und Kunst Baden- Wuerttemberg).}, GRID-CSIC
\footnote{GRID-CSIC (http://www.grid-csic.es), Grid infrastructure for
  advanced research at the Spanish National Research Council,
  Ref. 200450E494.}  project FIS2012-30634 (funded by the Spanish MEC). We also thank the Junta de Andaluc\'{i}a for its support via P11-FQM-7074 project (Spain).

\sloppy

\end{document}